\documentclass[
12pt,
preprint,preprintnumbers,nofootinbib,
groupedaddress,superscriptaddress,amsmath,amssymb]{revtex4}
\usepackage{graphicx}
\usepackage{dcolumn}
\usepackage{bm}
\usepackage{amssymb}
\usepackage{amsmath}
\usepackage{epsfig}    
\usepackage{color}
\usepackage{hhline}

\def\be{\begin{equation}}
\def\ee{\end{equation}}
\newcommand{\bea}{\begin{eqnarray}}
\newcommand{\eea}{\end{eqnarray}}
\newcommand{\nn}{\nonumber}

\numberwithin{equation}{section}

\begin{document}
{\begin{flushright}{KIAS-P22026, APCTP Pre2022 - 005}\end{flushright}}

\title{Natural mass hierarchy among three heavy Majorana neutrinos for resonant leptogenesis under modular $A_4$ symmetry}
%

\author{Dong Woo Kang}
\email{dongwookang@kias.re.kr}
\affiliation{School of Physics, KIAS, Seoul 02455, Korea}

\author{Jongkuk Kim}
\email{jkkim@kias.re.kr}
\affiliation{School of Physics, KIAS, Seoul 02455, Korea}

\author{Takaaki Nomura}
\email{nomura@scu.edu.cn}
\affiliation{College of Physics, Sichuan University, Chengdu 610065, China}

\author{Hiroshi Okada}
\email{hiroshi.okada@apctp.org}
\affiliation{Asia Pacific Center for Theoretical Physics (APCTP) - Headquarters San 31, Hyoja-dong,
Nam-gu, Pohang 790-784, Korea}
\affiliation{Department of Physics, Pohang University of Science and Technology, Pohang 37673, Republic of Korea}

\date{\today}

\begin{abstract}
It is clear that matter is dominant in the Universe compared to antimatter.
We call this problem baryon asymmetry. 
The baryon asymmetry is experimentally determined by both cosmic microwave background and  big bang nucleosynthesis measurements. 
To resolve the baryon number asymmetry of the Universe as well as neutrino oscillations, we study a radiative seesaw model in a modular $A_4$ symmetry.
Degenerate heavy Majorana neutrino masses can be naturally realized in an appropriate assignments under modular $A_4$  with large imaginary part of modulus $\tau$, and it can induce measured baryon number via resonant leptogenesis that is valid in around TeV scale energy theory.
We also find that the dominant contribution to the CP asymmetry arises from Re[$\tau$] through our numerical analysis satisfying the neutrino oscillation data.
\end{abstract}
\maketitle
\newpage

\section{Introduction}
{Understanding the origin of the observed imbalance in baryonic matter remains one of the important problems, so-called Baryon Asymmetry of the Universe (BAU) problem, in particle physics and cosmology.
The baryon abundance at present Universe obtained from Cosmic Microwave Background anisotropy and Big Bang Nucleosynthesis is \cite{Planck:2018vyg}
\begin{align}
	\Omega_B h^2 =0.0223 \pm 0.0002.
\end{align}
This is related to the baryon asymmetry which is given by
\begin{align}
	Y_B\equiv \frac{n_B}{s}\simeq 0.86\times10^{-10},\label{bau}
\end{align}
where $n_B$ is the number density of the baryon and $s$ is entropy density.

Theoretically resolving the BAU  is one of the important motivations to consider physics 
}
 beyond the Standard Model (SM).~\footnote{Even though there exists electroweak baryogenesis~\cite{Cohen:1993nk} within the context of SM, it would be almost ruled out due to requirement of too strong first order electroweak phase transition. {Non-thermal WIMP baryogenesis is one of the fascinating mechnisms since re-annihilation of dark matter provides the observed baryon asymmetry as well as the correct relic density \cite{Choi:2018kto}.}}
One attractive scenario is leptogenesis~\cite{Fukugita:1986hr} that is intimately related to smallness of neutrino masses through canonical seesaw mechanism~\cite{Minkowski:1977sc, Mohapatra:1979ia} by introducing heavy Majorana fermions.
The decay of the heavy Majorana neutrino creates lepton asymmetry and it is converted into baryon asymmetry by sphalerons around electroweak scale.
In order to generate observed baryon asymmetry in Eq.~(\ref{bau}), we need rather large mass  of the Majorana neutrinos that is $10^{10}$ GeV at least. It implies that such heavy fermions could not be produced by any current collider experiments. 

{On the contrary,} resonant leptogenesis~\cite{Pilaftsis:2003gt, Pilaftsis:1997jf} is one of the promising candidates to explain BAU at low energy scale {that could be around TeV scale being within the reach of current or near future experiments.}
In order to realize the resonant leptogenesis, we need two almost degenerate Majorana fermions at least.~\footnote{More concretely, the maximum enhancement may be achieved if the mass splitting between two Majorana fermions is comparable to the decay width of either Majorana fermions.}
Then, we have an enhancement of the $CP$ asymmetry parameter~\cite{Flanz:1996fb} that can generate enough BAU even at low energy scale. The next task would be how to realize the almost degenerate masses, even though most of the {endeavors} {set} such a situation by hand unless symmetries are introduced.

Recently, a modular flavor symmetry was proposed in Ref.~\cite{Feruglio:2017spp,deAdelhartToorop:2011re}.
The symmetry has additional quantum number that is called modular weight, and flavor structures of dimensionless couplings such as Yukawas
are uniquely determined once the modular weights are fixed. As a result, more predictive models are possible compared to traditional flavor models. In fact, {a plethora of scenarios have been studied} after this idea up to now, {especially applying modular $A_4$ symmetry}~\cite{Feruglio:2017spp, Criado:2018thu, Kobayashi:2018scp, Okada:2018yrn, Nomura:2019jxj, Okada:2019uoy, deAnda:2018ecu, Novichkov:2018yse, Nomura:2019yft, Okada:2019mjf,Ding:2019zxk, Nomura:2019lnr,Kobayashi:2019xvz,Asaka:2019vev,Zhang:2019ngf, Gui-JunDing:2019wap,Kobayashi:2019gtp,Nomura:2019xsb, Wang:2019xbo,Okada:2020dmb,Okada:2020rjb, Behera:2020lpd, Behera:2020sfe, Nomura:2020opk, Nomura:2020cog, Asaka:2020tmo, Okada:2020ukr, Nagao:2020snm, Okada:2020brs, Yao:2020qyy, Chen:2021zty, Kashav:2021zir, Okada:2021qdf, deMedeirosVarzielas:2021pug, Nomura:2021yjb, Hutauruk:2020xtk, Ding:2021eva, Nagao:2021rio, king, Okada:2021aoi, Nomura:2021pld, Kobayashi:2021pav, Dasgupta:2021ggp, Liu:2021gwa, Nomura:2022hxs, Otsuka:2022rak,Kobayashi:2021ajl, Chauhan:2022gkz, Kikuchi:2022pkd, Kobayashi:2022jvy, Gehrlein:2022nss, Almumin:2022rml,Kashav:2022kpk}.
\footnote{Here, we provide useful review references for beginners~\cite{Altarelli:2010gt, Ishimori:2010au, Ishimori:2012zz, Hernandez:2012ra, King:2013eh, King:2014nza, King:2017guk, Petcov:2017ggy, Kobayashi:2022moq}.}
More interestingly on the resonant leptogenesis, {\it two degenerate Majorana fermion masses are arisen at two fixed points of modulus $\tau=i,\ i\infty$, once we assign three Majorana fermions $N^c$ to be triplet under $A_4$ with $-1$ modular weight!}
These fixed points are statistically favored in the flux compactification of Type IIB string theory \cite{Kobayashi:2021pav}.
\footnote{Generally, there are three fixed points adding $\tau=e^{2\pi i/3}$. But this point does not give degenerated masses under the same asignment.}
Under the assignment, the Majorana mass matrix ($M_N$) is given by
\begin{align}
M_N=
{M_0}
\left[\begin{array}{ccc}
2y_{1} & -y_{3} & -y_{2} \\ 
-y_{3} & 2y_{2} & -y_{1} \\ 
-y_{2} & -y_{1} &2 y_{3} \\ 
\end{array}\right],\label{eq:mn}
 \end{align}
 where $Y^{(2)}_3\equiv (y_1,y_2,y_3)$ is 
 {$A_4$ triplet modular form with two modular weights appearing in Majorana mass term $\frac12 M_0 Y^{(2)}_3 \overline{ N^c} N$; concrete forms of modular forms are found in Appendix.}
In the limit of $\tau=i$, we have $(0,M_0,M_0)$ mass eigenvalues after the diagonalization of Eq.~(\ref{eq:mn}).
The deviation from the $\tau=i$ gives small {mass splitting} of the two massive Majorana fermions as well as the small mass eigenvalue for the massless Majorana fermion. {We naturally get the small mass splitting between two massive Majorana fermions but it is hard to realize resonant leptogenesis with one light (almost massless) Majorana fermion.
In the limit of $\tau=i\infty$, on the other hand, we have $(M_0,M_0,2M_0)$ mass eigenvalues, i.e. three heavy Majorana fermions with two degenerated one.}
And the deviation along the Im[$\tau$] direction from the $\tau=i\infty$ provides small {mass splitting} of the lightest two massive Majorana fermions,
while the deviation along the Re[$\tau$] direction provides the CP asymmetry.

In this paper, we apply the modular $A_4$ symmetry with rather large Im$[\tau]$ to a supersymmetric radiative seesaw model,
and discuss how to realize the neutrino oscillation data and BAU simultaneously~\cite{Kashiwase:2012xd}. 
Radiative seesaw model~\cite{Ma:2006km} is known as a promising candidate to explain the neutrino oscillation data at low energy scale, and connect the neutrinos and dark matter or new particles. Moreover, several interesting flavor physics such as lepton flavor violations (LFVs) potentially come into our discussion.
Due to almost two degenerate Majorana fermions, the resonant leptogenesis is naturally realized and proper field assignments for modular weight assure the radiative seesaw model instead of an ad-hoc symmetry such as $Z_2$ that is introduced by hand.

Our manuscript is organized as follows.
{In Sec.~\ref{sect2}, we give our concrete model set up under $A_4$ modular symmetry, and formulate mass matrices of Majorana heavy neutrinos, inert scalar bosons, and active neutrinos. Then, we explain how to compare the experimental values.
In Sec.~\ref{sect3}, we show how to realize the resonant leptogenesis in our model.
In Sect.~\ref{sect4}, we perform the $\chi^2$ numerical analysis, and show our results to satisfy all the experimental constraints.
 Finally, we conclude and discuss in Sec.~\ref{sec:conclusion}.}

\section{Model}
\label{sect2}
{In this section, we review a radiative seesaw model in a modular $A_4$ symmetry. }
We introduce modular $A_4$ symmetry and assign $\{1,1',1''\}$ for $({\hat L_{e}},{\hat L_{\mu}},{\hat L_{\tau}})$ and  $\{1,1'',1'\}$ for $(\hat e^c,\hat \mu^c,\hat \tau^c)$ in order to consider the mass eigenbasis of charged-lepton sector.
Their modular weights are assigned to be $\{-2,-2,0\}$ for each family.
Note here that notation $\hat f$ for a  field $f$ indicates matter  chiral superfield including boson and fermion,
otherwise $f$ denotes boson or fermion {with even $R$-parity}.
In addition, we introduce three electrically neutral superfields $\hat N^c$ as discussed in introduction.
Then, we add $\hat \eta_{1}$ with $-3$ modular weight whose {corresponding scalar field} is important to connect the active neutrinos and the Majorana fermions, while $\hat\eta_2$ is introduced only to cancel the gauge chiral anomaly and this field does not contribute to the neutrino sector.~\footnote{$\hat H_2$ is requested by obtaining the SM Higgs mass through $\mu$ term. But, we suppose $H_2$ does not contribute to our main discussion as well.} The field contents and their assignments are summarized in Table~\ref{tab:1}.
%
\begin{center} 
\begin{table}[tb]
\begin{tabular}{||c||c|c|c||c|c|c|c|c||}\hline
 &\multicolumn{8}{c||}{Matter super fields}  \\ \hline \hline
Fields  & ~$({\hat L_{e}},{\hat L_{\mu}},{\hat L_{\tau}})$~ & ~$(\hat e^c,\hat\mu^c,\hat\tau^c)$~ & ~$\hat N^c$~ & ~$\hat H_1$~ & ~$\hat H_2$~  & ~$\hat\eta_1$~  & ~$\hat\eta_2$~  & ~$\hat\chi$~  \\\hline 
 $SU(2)_L$ & $\bm{2}$     & $\bm{1}$  & $\bm{1}$ & $\bm{2}$ & $\bm{2}$& $\bm{2}$ & $\bm{2}$ & $\bm{1}$    \\\hline 
$U(1)_Y$ & $\frac12$  & $-1$ & $0$  & $\frac12$  & -$\frac12$   & $\frac12$  & -$\frac12$     & $0$   \\\hline
 $A_4$ & $\{1,1',1''\}$ & $\{1,1'',1'\}$ & $3$ & $1$ & $1$ & $1$ & $1$ & $1$ \\\hline
 $-k$ & $\{-2,-2,0\}$ & $\{-2,-2,0\}$ & $-1$ & $0$ & $0$& $-3$ & $-3$ & $-3$   \\\hline
\end{tabular}
\caption{Matter chiral superfields and their charge assignments under $SU(2)_L\times U(1)_Y\times A_4$, where $-k$ is the number of modular weight.}
\label{tab:1}
\end{table}
\end{center}

\subsection{Majorana mass matrix}
At first, we develop our discussion on the  mass matrix of Majorana fermion $N$ given in Eq.~(\ref{eq:mn}).
At nearby $\tau=i\infty$, $(y_1,y_2,y_3)$ is expanded by $p_\epsilon$ and given by $(1+12p_\epsilon,-6p_\epsilon^{1/3},-18p_\epsilon^{2/3})$, where $|p_\epsilon|\equiv |e^{2\pi i \tau}|\ll 1$ and $(y_1,y_2,y_3)$ is given in Appendix A.
Then, the Majorana mass matrix in Eq.~(\ref{eq:mn}) is rewritten by
\begin{align}
M_N&\approx
M_0
\left[\begin{array}{ccc}
2(1+12p_\epsilon) & 18 p_\epsilon^{2/3} & 6 p_\epsilon^{1/3} \\ 
18 p_\epsilon^{2/3} & -12p_\epsilon^{1/3} & -(1+12p_\epsilon) \\ 
6 p_\epsilon^{1/3} & -(1+12p_\epsilon) &-36p_\epsilon^{2/3} \\ 
\end{array}\right].
 \end{align}
It is diagonalized by a unitary matrix $U_N$ as $D \equiv U_N^* M_N U_N^\dag$, and their mass eigenstates are defined by $\psi$, where $N=U_N^\dag \psi$. Therefore, $|D|^2=U_N M_N^\dag M_N U_N^\dag$.
Approximately $U_N$ and $|D|$ are given by the following forms:
\begin{align}
U_N&\approx
 \left[\begin{array}{ccc}
-3\sqrt2p_\epsilon^{1/3}(1-6p_\epsilon^{1/3}) &-\frac1{\sqrt2}(1-3\epsilon^{1/3}+\frac92{ p_\epsilon^{2/3}} ) 
&\frac1{\sqrt2}(1+3\epsilon^{1/3}-\frac{63}2 { p_\epsilon^{2/3}} ) \\ 
-\sqrt2 p_\epsilon^{1/3}(1-2 p_\epsilon^{1/3}) &\frac1{\sqrt2}(1+3\epsilon^{1/3}-\frac{35}2 {p_\epsilon^{2/3}}) 
 & \frac1{\sqrt2}(1-3\epsilon^{1/3}+\frac{13}2 {p_\epsilon^{2/3}})  \\ 
1-10p_\epsilon^{1/3} &-2p_\epsilon^{1/3}(1-14p_\epsilon^{1/3})  &2p_\epsilon^{1/3}(2-7p_\epsilon^{1/3}) \\ 
\end{array}\right]^* + {\cal O}(p_\epsilon),\\
|D|&\approx M_0
{\rm diag.}\left[1-6p_\epsilon^{1/3}-18p_\epsilon^{2/3},\ 1+6p_\epsilon^{1/3}+42 p_\epsilon^{2/3}, \ 
2+24 p_\epsilon^{2/3}\right] + {\cal O}(p_\epsilon) .
\end{align}
Thus, one can straightforwardly find that $|D|=M_0 \, {\rm diag.}[1,1,2]$ in the limit of $\tau \to i\infty$.

\subsection{Inert scalar boson mass matrices}
Inert scalar boson components of $\hat\eta_{1,2}$ and $\hat\chi$ contribute to the neutrino mass matrix.
The relevant Lagrangian among these bosons is given via soft  SUSY-breaking terms as follows:
 \begin{align}
-{\cal L}_{\rm soft}&= m_A  H_2 \eta_1 \chi +m_B^2\chi^2+m^2_{\eta_1}|\eta_1|^2+m^2_{\chi} |\chi|^2\\
&
+m^2_{H_1} |H_1|^2+m^2_{H_2} |H_2|^2
+m^2_{\eta_2} |\eta_2|^2
+\mu^2_{BH} H_1 H_2 +\mu^2_{B\eta} \eta_1 \eta_2 +m'_A H_1 \eta_2 \chi +{\rm h.c.},\nn
\end{align}
where the terms in the first line of RHS directly contributes to the neutrino mass matrix and some of mass parameters include modular forms.
When the neutral components are written in terms of real and imaginary part as $\chi=(\chi_R+i\chi_I)/\sqrt2$ and {$\eta_{1}=(\eta^+_{1}, (\eta_{R_{1}}+i\eta_{I_{1}})/\sqrt2)^T$,} the mass squared matrices in basis of $(\eta_1,\chi)_{R,I}$ are given by
\begin{align}
m_R^2&=
 \left[\begin{array}{cc}
m^2_{\eta_1} & \frac{v_2 m_A}{\sqrt2} \\ 
\frac{v_2 m_A}{\sqrt2} & m^2_{\chi} +m_B^2  \\ 
\end{array}\right],\quad 
m_I^2=
 \left[\begin{array}{cc}
m^2_{\eta_1} & -\frac{v_2 m_A}{\sqrt2} \\ 
-\frac{v_2 m_A}{\sqrt2} & m^2_{\chi} - m_B^2 \\ 
\end{array}\right],\label{eq:inert-mass}
\end{align}
these are diagonalized by $D_{R,I}^2=O_{R,I} M^2_{R,I} O^T_{R,I}$, where $O_{R,I}$ is an orthogonal matrix;
\begin{align}
 \left[\begin{array}{c}
\eta_{1} \\ 
\chi \\ 
\end{array}\right]_{R,I}
=
 \left[\begin{array}{cc}
c_{R,I}& s_{R,I} \\ 
-s_{R,I} &c_{R,I}  \\ 
\end{array}\right]
 \left[\begin{array}{c}
\varphi_1 \\ 
\varphi_2 \\ 
\end{array}\right]_{R,I},\quad 
O_{R,I}=
 \left[\begin{array}{cc}
c_{R,I}& s_{R,I} \\ 
-s_{R,I} &c_{R,I}  \\ 
\end{array}\right].
\end{align}
Here $\varphi_{1R,2R,1I,2I}$ are mass eigenvectors, $s_{R,I},\ c_{R,I}$ are respectively short-hand notations for $\sin\theta_{R,I}$, $\cos\theta_{R,I}$ which are given by a function of mass parameters in Eq.~(\ref{eq:inert-mass}).

\subsection{Neutrino sector}
\label{subsec:neutrino}
Now we can discuss the neutrino sector estimating neutrino mass at one-loop level.
Our valid renormalizable Lagrangian in terms of mass eigenstates of heavier Majorana fermions and inert scalar bosons, coming from superpotential, is given by
\begin{align}
-{\cal L}^\nu &=
\frac{1}{\sqrt2} \overline{\psi}_i (U_N)_{ia} y_{\eta_{ab}} \nu_b (c_R \varphi_{1R} +s_R \varphi_{2R})
+\frac{i}{\sqrt2} \overline{\psi}_i (U_N)_{ia} y_{\eta_{ab}} \nu_b (c_I \varphi_{1I} +s_I \varphi_{2I})+{\rm h.c.},\label{eq:neut}\\
y_\eta&=
\left[\begin{array}{ccc}
a_\eta & 0 & 0 \\ 
0 & b_\eta & 0  \\ 
0 & 0 & c_\eta  \\ 
\end{array}\right]
\left[\begin{array}{ccc}
y^{(6)}_1+ \epsilon_e y^{'(6)}_1 & y^{(6)}_3+ \epsilon_e y^{'(6)}_3 & y^{(6)}_2+ \epsilon_e y^{'(6)}_2 \\ 
y^{(6)}_3+ \epsilon_\mu y^{'(6)}_3 & y^{(6)}_2+ \epsilon_\mu y^{'(6)}_2 & y^{(6)}_1+ \epsilon_\mu y^{'(6)}_1  \\ 
y^{(4)}_2 & y^{(4)}_1 & y^{(4)}_3  \\ 
\end{array}\right],
\end{align}
where $a_{\eta},\ b_\eta,\ c_\eta$ are real without loss of generality after phase redefinition, while $\epsilon_{e,\mu}$ are complex values which would also be sources of CP asymmetry in addition to $\tau$. {Explicit forms of}
$(y^{(4)}_1,y^{(4)}_2,y^{(4)}_3)$, $(y^{(6)}_1,y^{(6)}_2,y^{(6)}_3)$, and $(y^{'(6)}_1,y^{'(6)}_2,y^{'(6)}_3)$ are given in Appendix A.
Then, the neutrino mass matrix is given as follows:
\begin{align}
(m_\nu)_{ab}
&=
-\frac{1}{2(4\pi)^2} (y_\eta)^T_{a\beta} (U^T_N)_{\beta i} D_{i} (U_N)_{i\alpha} (y_\eta)_{\alpha b}\nn\\
&\times
\left[
c^2_R F(m_{\varphi_{1R}},D_{i}) - c^2_I F(m_{\varphi_{1I}},D_{i})
+
s^2_R F(m_{\varphi_{2R}},D_{i}) - s^2_I F(m_{\varphi_{2I}},D_{i})
\right],\label{eq:neut1}
\end{align}
where
\begin{align}
 F(m_a,m_b)
 &=\frac{\ln \left(\frac{m_a^2}{m_b^2}\right)}{\frac{m_a^2}{m_b^2}-1}-1.
\end{align}
The neutrino mass matrix $m_\nu$ is then diagonalized by an unitary matrix $U_{\rm PMNS}$; $U_{\rm PMNS}^T m_\nu U_{\rm PMNS}\equiv {\rm diag}(m_1,m_2,m_3)$.  
We write two mass squared differences measured by the experiments, 
\begin{align}
&\Delta m_{\rm sol}^2=m_2^2 - m_1^2,\\
&(\mathrm{NH}):\  \Delta m_{\rm atm}^2 =m_3^2 - m_1^2,
\quad
(\mathrm{IH}):\    \Delta m_{\rm atm}^2 =m_2^2 - m_3^2,
 \end{align}
where $\Delta m_{\rm sol}^2$ is solar mass squared difference and  $\Delta m_{\rm atm}^2$ is atmospheric neutrino mass squared difference whose form depends on
{whether neutrino mass ordering is} normal hierarchy (NH) or inverted hierarchy (IH). 
 %
$U_{\rm PMNS}$ is parametrized by three mixing angle $\theta_{ij} (i,j=1,2,3; i < j)$, one CP violating Dirac phase $\delta_{CP}$,
and two Majorana phases $\{\alpha_{21}, \alpha_{32}\}$ as follows:
\begin{equation}
U_{\rm PMNS}= 
\begin{pmatrix} c_{12} c_{13} & s_{12} c_{13} & s_{13} e^{-i \delta_{CP}} \\ 
-s_{12} c_{23} - c_{12} s_{23} s_{13} e^{i \delta_{CP}} & c_{12} c_{23} - s_{12} s_{23} s_{13} e^{i \delta_{CP}} & s_{23} c_{13} \\
s_{12} s_{23} - c_{12} c_{23} s_{13} e^{i \delta_{CP}} & -c_{12} s_{23} - s_{12} c_{23} s_{13} e^{i \delta_{CP}} & c_{23} c_{13} 
\end{pmatrix}
\begin{pmatrix} 1 & 0 & 0 \\ 0 & e^{i \frac{\alpha_{21}}{2}} & 0 \\ 0 & 0 & e^{i \frac{\alpha_{31}}{2}} \end{pmatrix},
\end{equation}
where $c_{ij}$ and $s_{ij}$ stands for $\cos \theta_{ij}$ and $\sin \theta_{ij}$ respectively. 
Then, {each mixing angle} is given in terms of the component of $U_{\mathrm{PMNS}}$ as follows:
\begin{align}
\sin^2\theta_{13}=|(U_{\mathrm{PMNS}})_{13}|^2,\quad 
\sin^2\theta_{23}=\frac{|(U_{\mathrm{PMNS}})_{23}|^2}{1-|(U_{\mathrm{PMNS}})_{13}|^2},\quad 
\sin^2\theta_{12}=\frac{|(U_{\mathrm{PMNS}})_{12}|^2}{1-|(U_{\mathrm{PMNS}})_{13}|^2}.
\end{align}
Also, we compute the Jarlskog invariant, $\delta_{CP}$ derived from PMNS matrix elements $U_{\alpha i}$:
\begin{equation}
J_{CP} = \text{Im} [U_{e1} U_{\mu 2} U_{e 2}^* U_{\mu 1}^*] = s_{23} c_{23} s_{12} c_{12} s_{13} c^2_{13} \sin \delta_{CP},
\end{equation}
and the Majorana phases are also estimated in terms of other invariants $I_1$ and $I_2$:
\begin{equation}
I_1 = \text{Im}[U^*_{e1} U_{e2}] = c_{12} s_{12} c_{13}^2 \sin \left( \frac{\alpha_{21}}{2} \right), \
I_2 = \text{Im}[U^*_{e1} U_{e3}] = c_{12} s_{13} c_{13} \sin \left( \frac{\alpha_{31}}{2} - \delta_{CP} \right).
\end{equation}
Additionally, the effective mass for the neutrinoless double beta decay is given by
\begin{align}
\langle m_{ee}\rangle=  |m_1 \cos^2\theta_{12} \cos^2\theta_{13}+m_2 \sin^2\theta_{12} \cos^2\theta_{13}e^{i\alpha_{21}}+m_3 \sin^2\theta_{13}e^{i(\alpha_{31}-2\delta_{CP})}|,
\end{align}
where its observed value could be measured by KamLAND-Zen in future~\cite{KamLAND-Zen:2016pfg}. 
We will adopt the neutrino experimental data in NuFit5.0~\cite{Esteban:2018azc} in order to perform the numerical $\chi^2$ analysis.

\section{Resonant leptogenesis}
\label{sect3}
{Now we consider non-thermal leptogenesis where population of the lightest sterile neutrino increases as the temperature decreases, so-called freeze-in production. 
The generation of the lepton asymmetry is non-equilibrium decay process of the lightest neutrino $\psi_1$. 
The dominant contribution to the $CP$ asymmetry is arisen from the interference between tree and one-loop {diagrams for} decays of $\psi_1$ via Yukawa coupling $y_\eta$.} 
In general, there are two one-loop decay modes; vertex correction diagram and self-energy correction, but 
the self energy correction is dominant in our case since $\psi_1$ and $\psi_2$ have degenerate mass.
Hence, the main source of CP asymmetry $\epsilon_1$ is approximately given as follows:
\begin{align}
&\epsilon_1\approx \frac{{\rm Im}(h^\dag h)^2_{12}}{(h^\dag h)^2_{11}(h^\dag h)^2_{22}}
\frac{(D^2_{1}-D^2_{2})D_{1} \Gamma_{2}}
{(D^2_{1}-D^2_{2})^2+ D^2_{1} \Gamma_{\psi_2}^2},\\
&\Gamma_{\psi_2} = \frac{|h_{2i}|^2}{4\pi}D_{2} \left(1-\frac{m_{\eta_1}^2}{D_{2}^2}\right)^2,
\end{align}
where $h\equiv U_N y_\eta$, and we expect $m_{\eta_1}\approx m_{\varphi_{1R}}\approx m_{\varphi_{1I}}$.
Then, we obtain the lepton asymmetry by solving the {approximated} Boltzmann equations {for $\psi_1$ and lepton number densities} as follows:
 \begin{align}
&\frac{dY_{\psi_1}}{dz}\approx -\frac{z}{sH(D_{1})} \left(\frac{Y_{\psi_1}}{Y_{\psi_1}^{eq}}-1\right)\gamma_D^{\psi_1},\\
&\frac{dY_{L}}{dz}\approx \frac{z}{sH(D_{1})} \left(\frac{Y_{\psi_1}}{Y_{\psi_1}^{eq}}-1\right)\epsilon_1 \gamma_D^{\psi_1},\\
&\gamma_D^{\psi_1} = \sum_{i=1}^3 \frac{|h_{1i}|^2}{4\pi^3} D_1^4 \left(1-\frac{m_{\eta_1}^2}{D_1^2}\right)^2
\frac{K_1(z)}{z},
\end{align}
{where $z\equiv D_1/T$, $T$ being temperature, $H(D_1)=1.66 {g_*^{1/2}} D_1^2/M_{pl}$, $g_*\approx 100$ is the number of relativistic degrees of freedom, and Plank mass $M_{pl}\approx 1.2\times 10^{19}$ GeV.
We denote the entropy dentsity as $Y_{\psi_1}\equiv n_{\psi_1}/s$ and {$Y_L\equiv( n_{L}-n_{\bar{L}})/s$}, $s=2\pi^2 g_* T^3/45$. }
Furthermore, $Y_{\psi_1}^{eq}$ is given by $45 z^2 K_2(z)/(2\pi^4 g_*)$, $K_{1(2)}(z)$, where the modified Bessel function of the first(second) kind.  
Once the lepton asymmetry $Y_L$ is generated, it is converted by $B+L$ violating spharelon transitions~\cite{Klinkhamer:1984di, Kuzmin:1985mm}. Then, the conversion rate is straightforwardly derived by the chemical equilibrium conditions and it is found as 
\begin{align}
Y_B=-\frac{8}{23} Y_{L}(z_{EW}),
\end{align}
where $z_{EW}$ corresponds to the spharelon decoupling temperature and we set to be $T_{EW}=100$ GeV.
%
Here, we should mention scattering processes that should be included in the above Boltzmann equations in general, since these processes also change the number of  lepton  as well as $\psi$.
The lepton number is changed by the processes of $\ell\ell\to\eta\eta$ and $\eta\ell\to\eta^\dag\bar\ell$, and the change of $\psi_1$ number is caused by $\psi_1\psi_1\to\ell\bar\ell$ and $\psi_1\psi_1\to\eta\eta^\dag$~\cite{Kashiwase:2012xd}.
Even though there exist resonant enhancement, the wash out processes from these scattering processes could be dominant depending on the parameter region. In order to suppress these processes enough, we should work on, e.g., $D_1\lesssim 2.5$ TeV when $|h|\approx10^{-4}$~\footnote{The order $10^{-4}$ comes from a prior research in ref.~\cite{Kashiwase:2012xd}. In this situation, LFVs are suppressed enough to satisfy the current experiments. In fact, the stringent constraint of $\mu\to e\gamma$ gives $4.2\times 10^{-13}$ of the branching ratio, while our maximum value of this process is $10^{-16}$ at most in our numerical analysis. Thus, we do not mention LFVs furthermore.}. Otherwise, we have to include the scattering processes.
{In our numerical analysis, we  compute the simplified Boltzmann equations by carefully checking whether {the washout} processes can be neglected or not
for our allowed parameter sets from neutrino data fitting.
In addition, the number density of $N_1$ is generated by freeze-in mechanism and it becomes the same order as that of thermal equilibrium for temperature higher than electroweak scale.
 }

\section{Numerical analysis}
\label{sect4}

{In this section, we carry out numerical analysis searching for allowed parameter region satisfying phenomenological constraints.}
{\it Note first that we find there is no allowed regions simultaneously satisfying the observed neutrino oscillation data and BAU in case of IH. Thus, we focus on the case of NH only.}
We show the allowed region with $\chi^2$ analysis in the lepton sector, where we randomly select the values of input parameters within the following ranges,
\begin{align}
&|{\rm Re}(\tau)|  \in [10^{-3},0.5],\ {\rm Im}(\tau) \in [2,10],\ \{ a_\eta, b_\eta, c_\eta\}   \in [10^{-8},10^{-3}],
\ \{ |\epsilon_e|, |\epsilon_\mu|\}   \in [10^{-3},10^{3}] \nonumber \\
& M_0 \in [10^2,10^5] \, {\rm GeV}, \  m_{\eta_1} \in [70,10^3] \, {\rm GeV}, \  m_{\chi} \in [1, 10^3] \, {\rm GeV}, \  \{m_{A},m_{B}\} \in [0.001, 50] \, {\rm GeV},
\end{align}
where we expect the source of CP asymmetry arises from three complex values $\tau,\ \epsilon_e,\ \epsilon_\mu $. 
{We also require inert scalar bosons are lighter than $\psi_1$ so that $\psi_1$ can decay into scalar boson and lepton to realize leptogenesis.}
Then, we accumulate the data if five measured neutrino oscillation data; ($\Delta m^2_{\rm atm},\ \Delta m^2_{\rm sol},\ \sin^2\theta_{13},\ \sin^2\theta_{23},\ \sin^2\theta_{12}$)~\cite{Esteban:2018azc} and BAU in Eq.~(\ref{bau}) are satisfied at the same time. Notice here that we regard $\delta_{CP}$ as a predicted value due to large ambiguity of experimental result in $3\sigma$ interval.

\begin{figure}[tb]\begin{center}
\includegraphics[width=80mm]{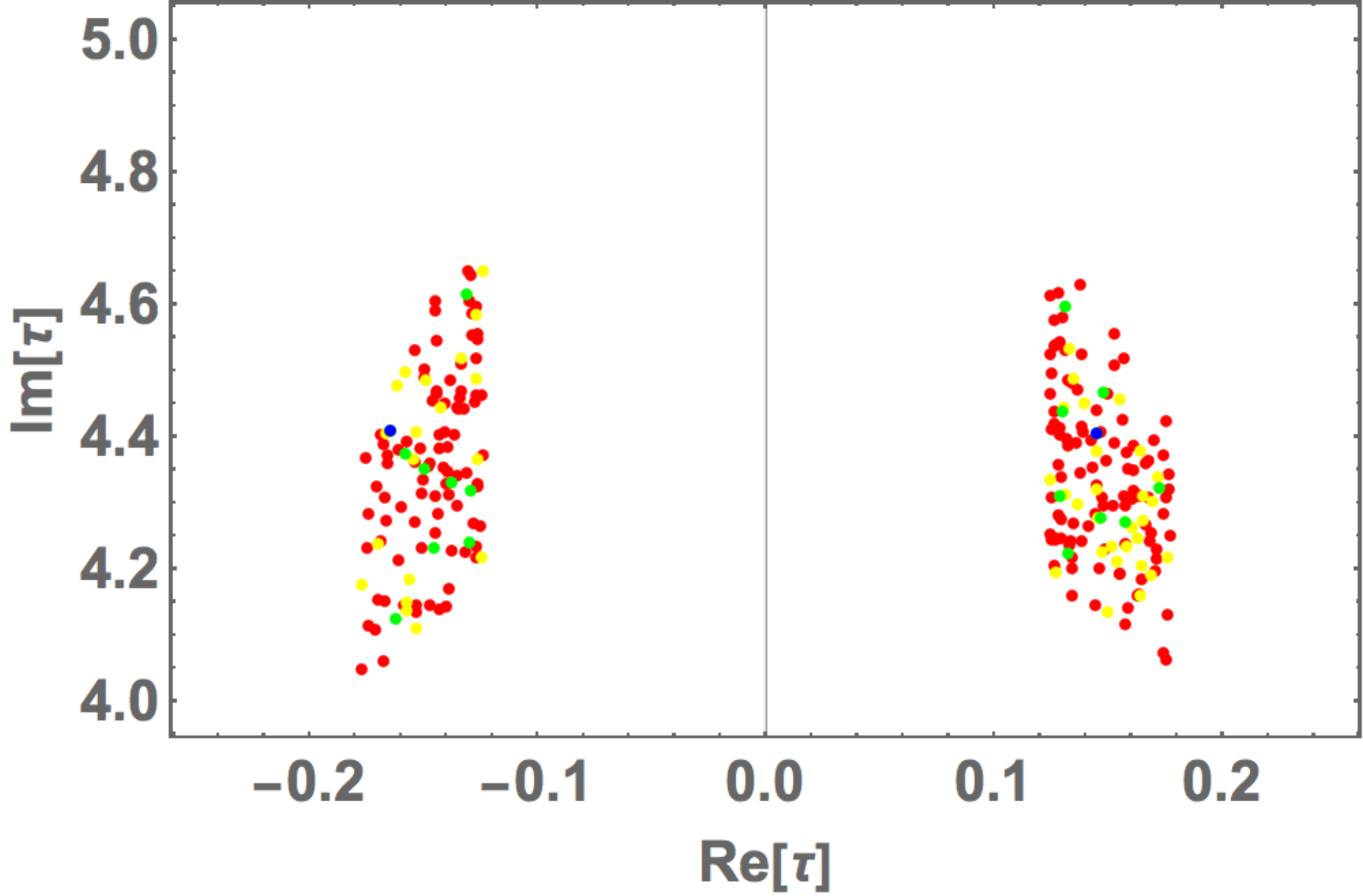}
\includegraphics[width=80mm]{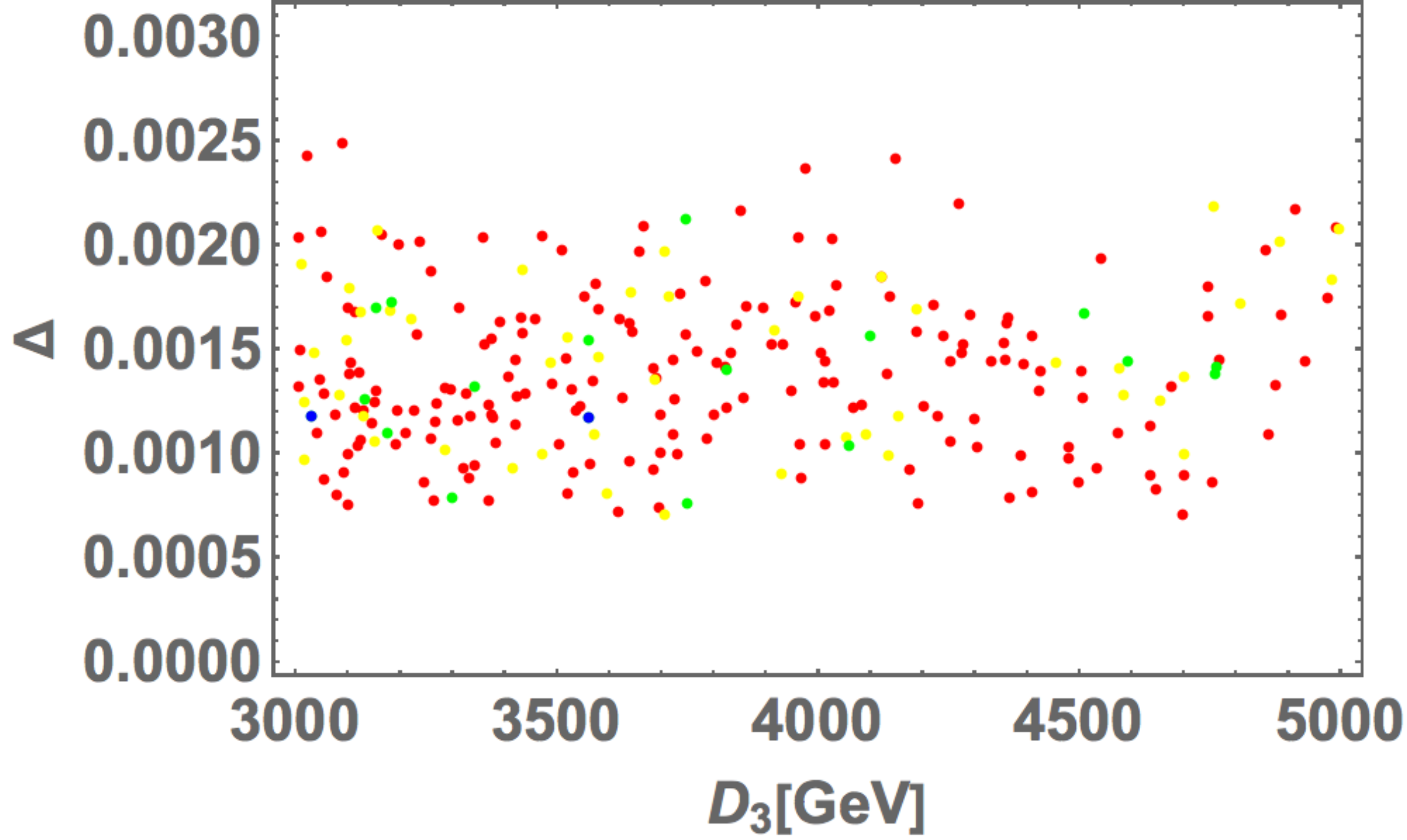}
\caption{The left plot shows allowed region of $\tau$, while the right one shows the mass degeneracy of $D_1$ and $D_2$ that is defined by $\Delta \equiv (D_2-D_1)/D_2$ in terms of $D_3$ in GeV unit, where $D_2(\sim D_1)\sim D_3/2$.
{Each point in blue, green, yellow, and red color} represents the allowed region in $\sigma\le1$, $1<\sigma\le2$, $2<\sigma\le3$, and $3<\sigma\le5$ interval, respectively. }   
\label{fig:tau1_nh}\end{center}\end{figure}
The left plot of Fig.~\ref{fig:tau1_nh} shows allowed region of $\tau$, while the right one {shows} the mass degeneracy of $D_1$ and $D_2$ that is defined by $\Delta \equiv (D_2-D_1)/D_2$ in terms of $D_3$ in GeV unit, where $D_2(\sim D_1)\sim D_3/2$.
{Each point in blue, green, yellow, and red color} represents the allowed region in $\sigma\le1$, $1<\sigma\le2$, $2<\sigma\le3$, and $3<\sigma\le5$ interval, respectively. 
These figures suggest that  $0.12\lesssim {\rm Re}|\tau|\lesssim0.18$ and  $4.05\lesssim {\rm Im}[\tau]\lesssim4.65$,
$7\times10^{-4}\lesssim\Delta\lesssim2.5\times10^{-3}$, and $3000 \ {\rm GeV}\lesssim D_3\lesssim5000 \ {\rm GeV}$.
Note here that the {smaller} $\Delta$ directly corresponds to the larger imaginary part of $\tau$, while the $ {\rm Re}[\tau]$ plays a role in generating one of the CP sources together with $\epsilon_{e,\mu}$. 
{We thus find sizable Re$[\tau]$ value is required in our allowed region and}
 it suggests that {\it the main source of CP asymmetry would arise from $\tau$.} 

\begin{figure}[tb]\begin{center}
\includegraphics[width=80mm]{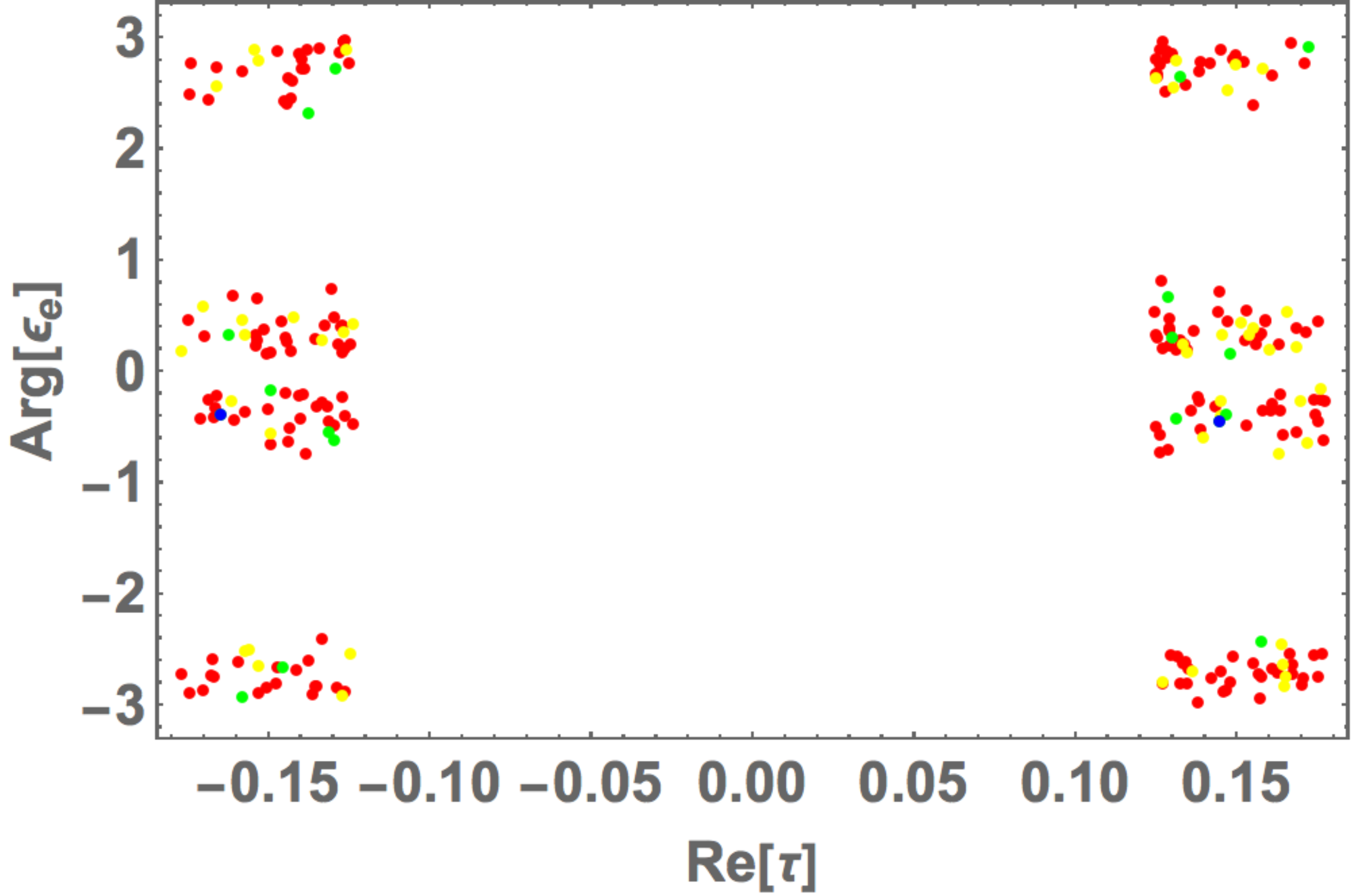}
\includegraphics[width=80mm]{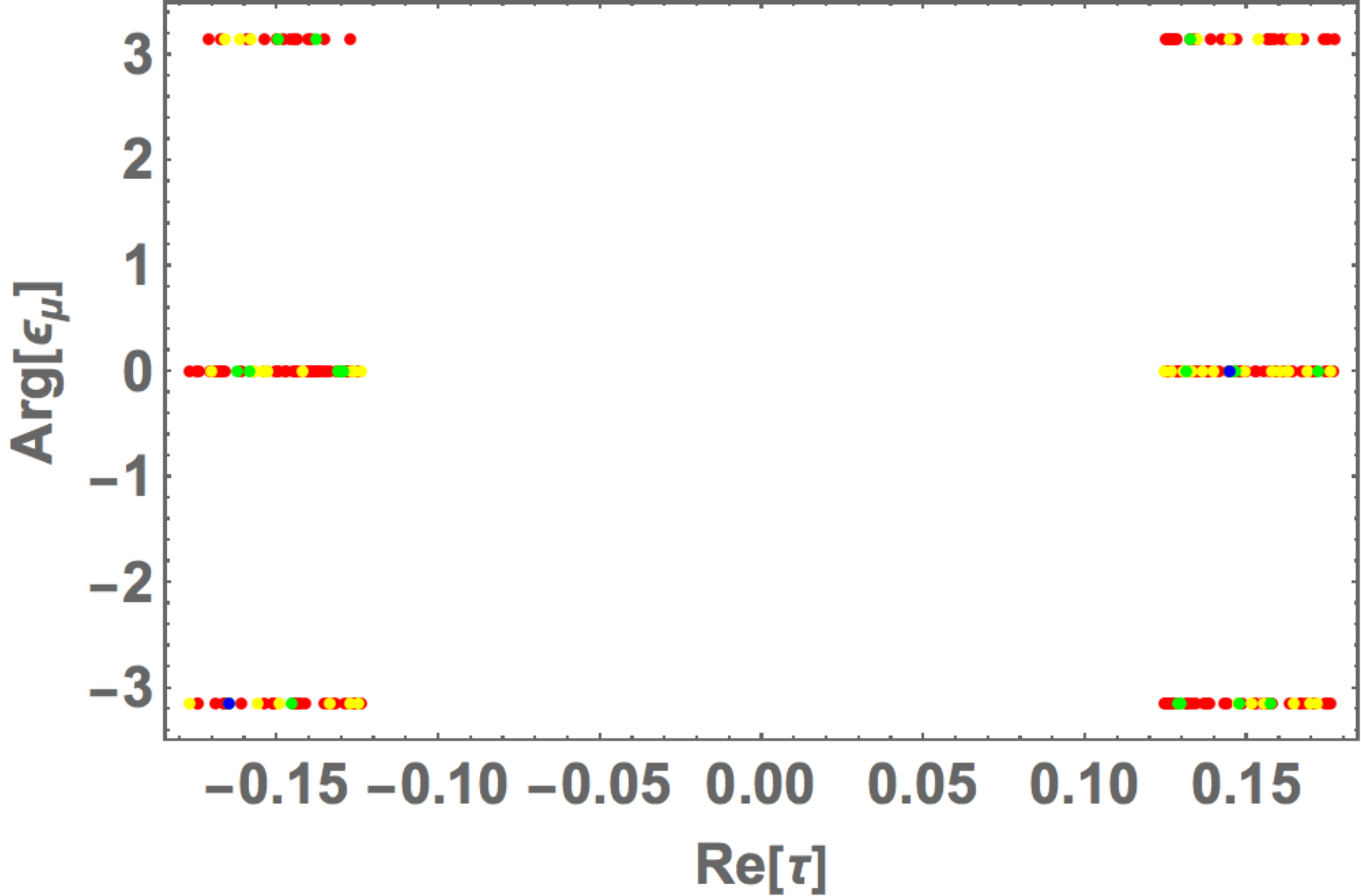}
\caption{The left(right) plot shows allowed region of ${\rm Arg}[\epsilon_{e(\mu)}]$ in terms of $ {\rm Re}[\tau]$, where the color legends are the same as the one in Fig.~\ref{fig:tau1_nh}. }   
\label{fig:cp-source}\end{center}\end{figure}
In order to investigate the main source of CP asymmetry, we demonstrate the plots of arguments of $\epsilon_{e,\mu}$ in terms of $ {\rm Re}[\tau]$ in Fig.~\ref{fig:cp-source}, where the color legends are the same as the one in Fig.~\ref{fig:tau1_nh}.
Remarkably, argument of $\epsilon_\mu$ is almost zero while the one of $\epsilon_e$ is slightly deviated from zero.
It comes from the resonant leptogenesis and neutrino oscillation.
In order to achieve the resonant leptogenesis, we need $(y_1,y_2,y_3)\sim(1,0,0)$.
It implies that $(y_1^{(6)},y_2^{(6)},y_3^{(6)})\sim(1,0,0)$ and $(y_1^{'(6)},y_2^{'(6)},y_3^{'(6)})\sim(0,0,0)$,
leading $y_\eta$ to
\begin{align}
y_\eta&\sim
\left[\begin{array}{ccc}
1 & 0 & 0 \\ 
0 & 0 & 1  \\ 
0 & 1 & 0  \\ 
\end{array}\right].
\end{align}
On the other hand, $y_{\eta_{12,13}} \ll y_{\eta_{21,32}}$ are required in order to obtain the observable neutrino oscillation data.
In order to compensate the tiny components of $Y^{(6)}_3,Y^{(6)}_{3'}$, rather large value of ${\rm Re}[\epsilon_\mu]$ is needed.
Thus, the argument of $\epsilon_\mu$ is relatively tiny.
Since both the arguments of  $\epsilon_{e,\mu}$ are localized at nearby $0$ and $\pi$,
{\it  the main origin of CP asymmetry indeed originates from $\tau$.}

\begin{figure}[tb]\begin{center}
\includegraphics[width=80mm]{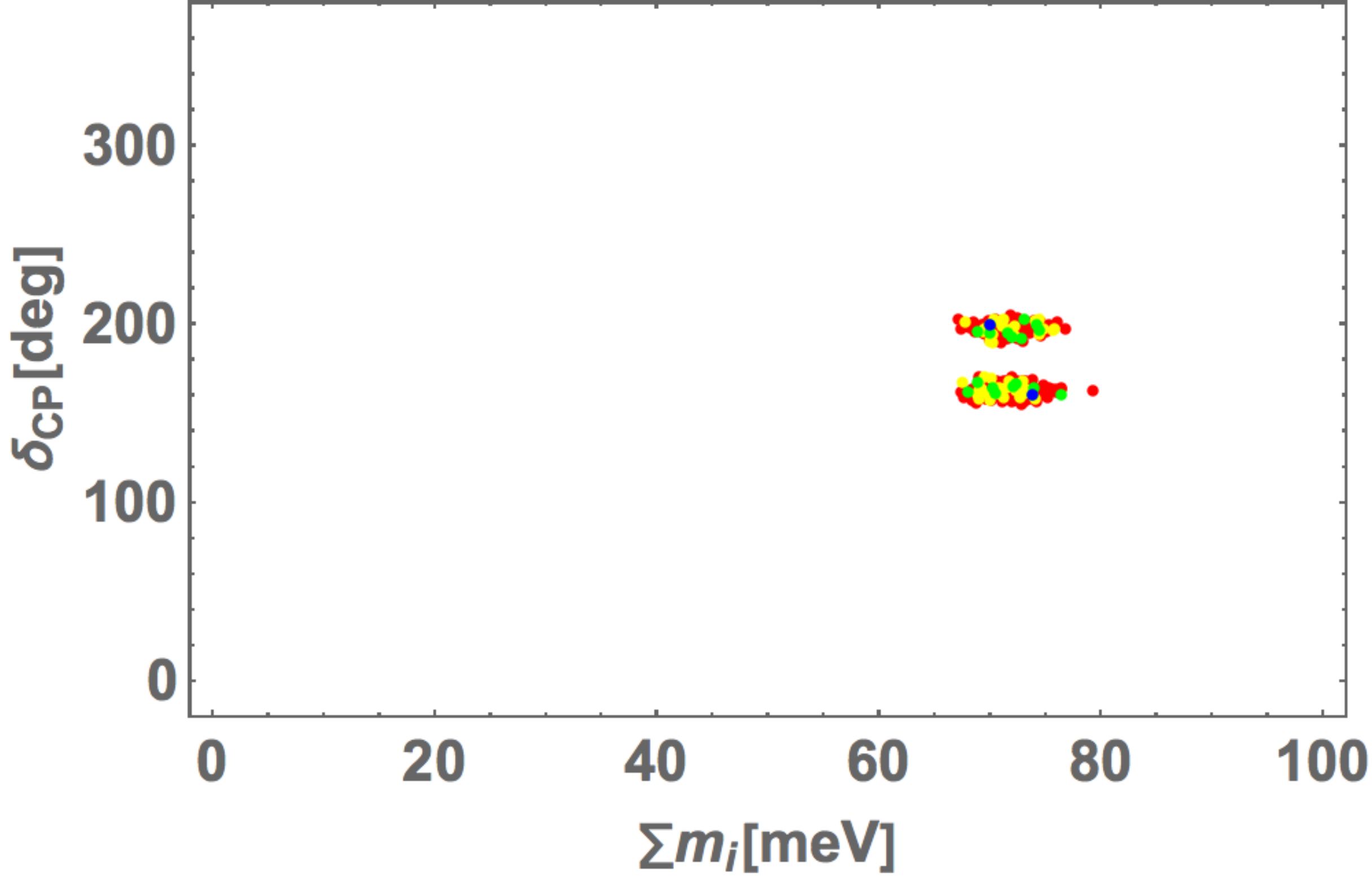}
\caption{ Allowed region of $\delta_{\rm CP}$ in terms of $\sum m_i$,
where the color legends are the same as the one in Fig.~\ref{fig:tau1_nh}.}   
\label{fig:sumdcp}\end{center}\end{figure}
The Fig.~\ref{fig:sumdcp} shows the allowed region of $\delta_{\rm CP}$ in terms of $\sum m_i$,
where the color legends are the same as the one in Fig.~\ref{fig:tau1_nh}.
We find two small localized regions of
$150^\circ \lesssim \delta_{\rm CP}\lesssim 170^\circ$ and $180^\circ \lesssim \delta_{\rm CP}\lesssim 200^\circ$ with $65\ {\rm meV}\lesssim \sum m_i \lesssim 80\ {\rm meV}$,
where the up island corresponds to the region of ${\rm Re}[\tau]<0$ while the  down one corresponds to the region of ${\rm Re}[\tau]>0$.
{\it Our result would favor  the best fit value of $\delta_{\rm CP}=195^\circ$.}

\begin{figure}[tb]\begin{center}
\includegraphics[width=80mm]{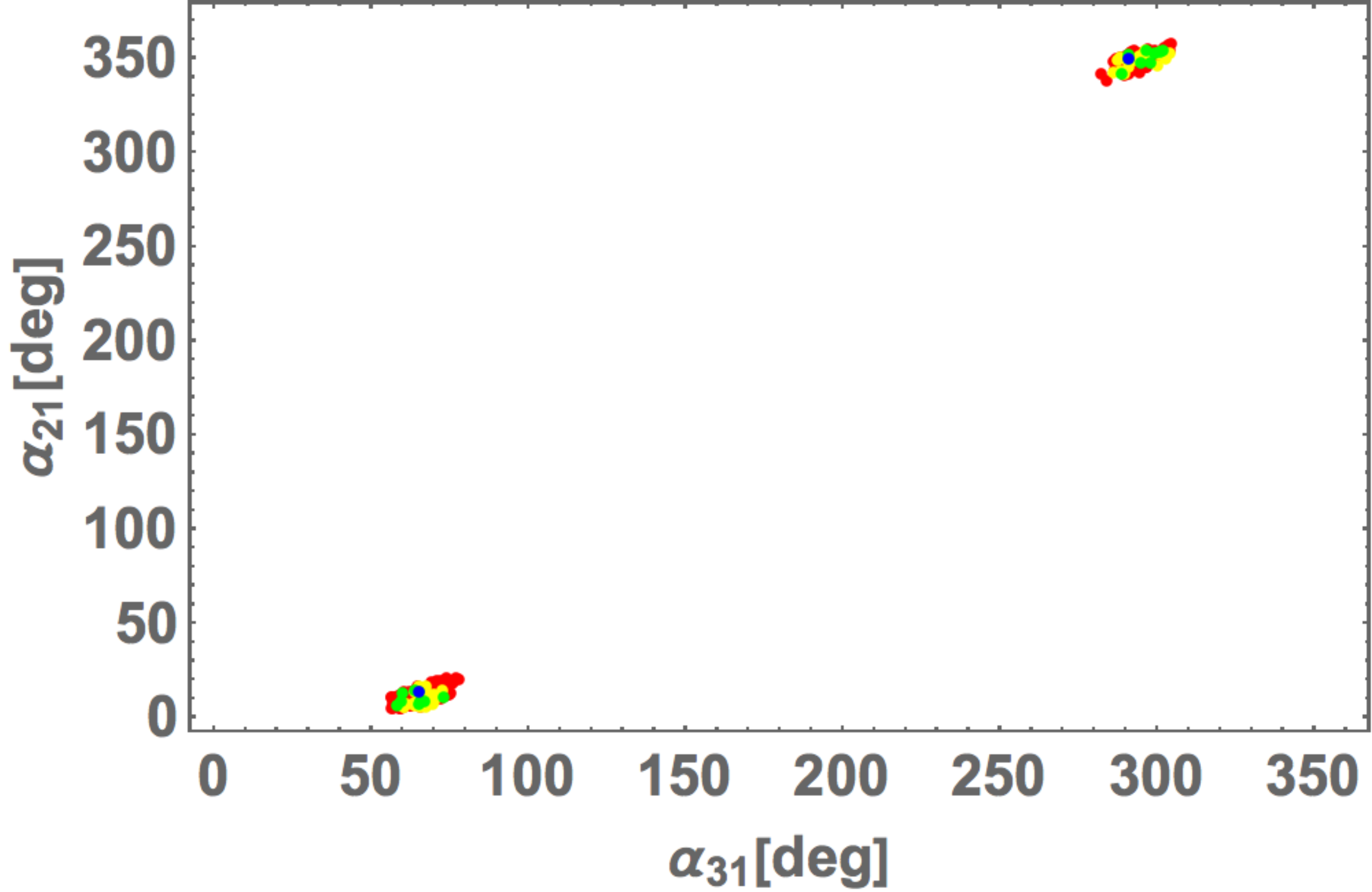}
\caption{Allowed region of Majorana phases $\alpha_{21}$ and $\alpha_{31}$, where the color legends are the same as the one in Fig.~\ref{fig:tau1_nh}.}   
\label{fig:majo}\end{center}\end{figure}
The Fig.~\ref{fig:majo} shows Majorana phases $\alpha_{21}$ and $\alpha_{31}$, where the color legends are the same as the one in Fig.~\ref{fig:tau1_nh}.
We have two localized islands of  $10^\circ\lesssim\alpha_{21}\lesssim30^\circ,\ 55^\circ\lesssim\alpha_{31}\lesssim80^\circ$,
 and $330^\circ\lesssim\alpha_{21}\lesssim360^\circ,\ 280^\circ\lesssim\alpha_{31}\lesssim310^\circ$,
where the up-right island corresponds to the region of ${\rm Re}[\tau]<0$ while the  down-left one corresponds to the region of ${\rm Re}[\tau]>0$.

\begin{figure}[tb]\begin{center}
\includegraphics[width=80mm]{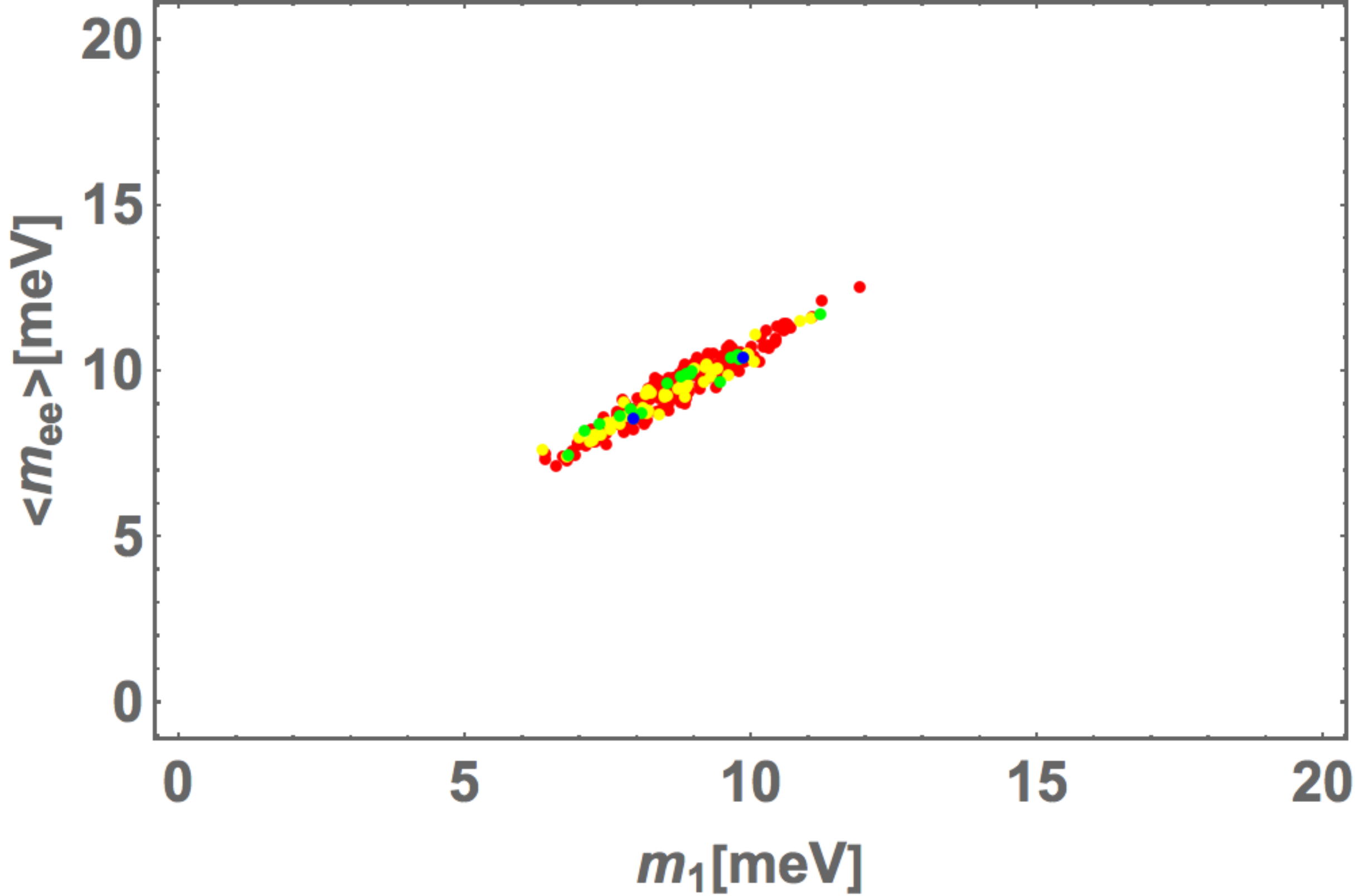}
\caption{Allowed region of $\langle m_{ee}\rangle$ in terms of the lightest neutrino mass $m_1$ meV, where the color legends are the same as the one in Fig.~\ref{fig:tau1_nh}.}   
\label{fig:m1_mee}\end{center}\end{figure}
The Fig.~\ref{fig:m1_mee} shows $\langle m_{ee}\rangle$ in terms of the lightest neutrino mass $m_1$ meV, where the color legends are the same as the one in Fig.~\ref{fig:tau1_nh}.
We obtain $7 \ {\rm meV}\lesssim \langle m_{ee}\rangle\lesssim12 \ {\rm meV}$
 and $6 \ {\rm meV}\lesssim  m_{1}\lesssim13 \ {\rm meV}$.

\begin{figure}[tb]\begin{center}
\includegraphics[width=80mm]{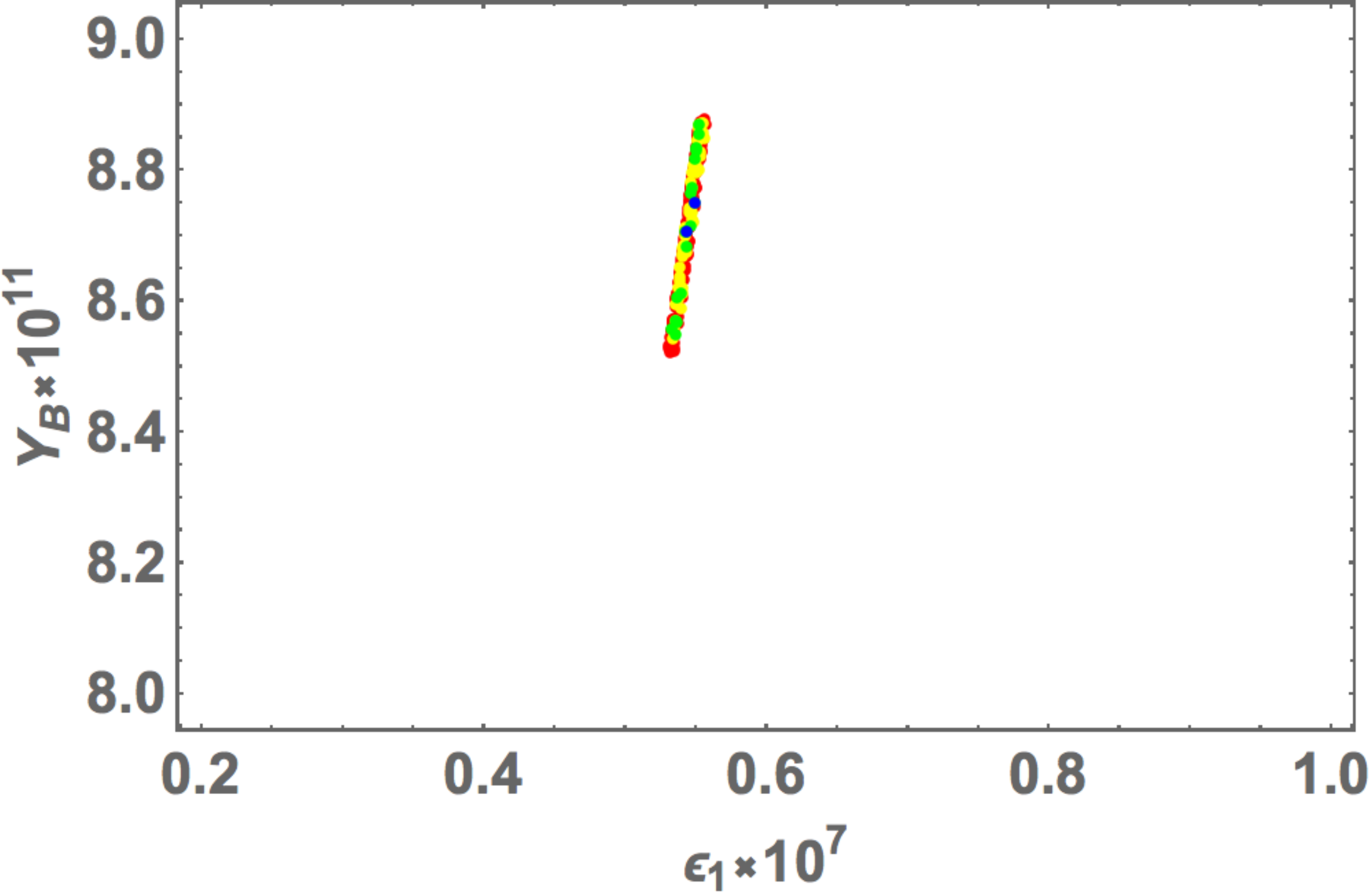}
\caption{Allowed region of the baryon asymmetry $Y_B$ in terms of the CP asymmetry source $\epsilon_1$, where the color legends are the same as the one in Fig.~\ref{fig:tau1_nh} and the $Y_B$ is taken within the observed range in Eq.(\ref{bau}).
}   
\label{fig:e1-yb}\end{center}\end{figure}
We show the baryon asymmetry $Y_B$ in terms of the CP asymmetry source $\epsilon_1$ in Fig.~\ref{fig:e1-yb}, where the color legends are the same as the one in Fig.~\ref{fig:tau1_nh} and the $Y_B$ is taken within the observed range in Eq.(\ref{bau}).
Here, $\epsilon_1$ is about $5\times 10^{-8}$ that is favored by the prior analysis in Ref.~\cite{Kashiwase:2012xd}.
Finally, we show a benchmark point for NH in Table~\ref{bp-tab_nh} that provide minimum $\sqrt{\chi^2}$ in our numerical analysis.

\begin{table}[h]
	\centering
	\begin{tabular}{|c|c|c|} \hline 
			\rule[14pt]{0pt}{0pt}
 		&  NH  \\  \hline
			\rule[14pt]{0pt}{0pt}
		$\tau$ & $-0.165 + 4.41 i$       \\ \hline
		\rule[14pt]{0pt}{0pt}
%
		$[a_\eta, b_\eta,c_\eta]\times 10^5$ & $[-4.94, -8.42, 6.71]$   \\ \hline
		\rule[14pt]{0pt}{0pt}
		$[M_0, M_{\eta_1}, m_\chi]/{\rm GeV}$ & $[1780, 199, 70.9]$     \\ \hline
		\rule[14pt]{0pt}{0pt}
		$[(v_2 m_A/\sqrt2)^{1/2}, m_B]/{\rm GeV}$ & $[45.1, 0.0324]$     \\ \hline
		\rule[14pt]{0pt}{0pt}
		$[\epsilon_e, \epsilon_\mu] $ & $[24.0 - 9.80 i, -761 - 0.00522 i]$    \\ \hline
		\rule[14pt]{0pt}{0pt}
		$\Delta m^2_{\rm atm}$  &  $2.47\times10^{-3} {\rm eV}^2$   \\ \hline
		\rule[14pt]{0pt}{0pt}
		$\Delta m^2_{\rm sol}$  &  $7.37\times10^{-5} {\rm eV}^2$        \\ \hline
		\rule[14pt]{0pt}{0pt}
		$\sin\theta_{12}$ & $ 0.544$   \\ \hline
		\rule[14pt]{0pt}{0pt}
		$\sin\theta_{23}$ &  $ 0.743$   \\ \hline
		\rule[14pt]{0pt}{0pt}
		$\sin\theta_{13}$ &  $ 0.151$   \\ \hline
		\rule[14pt]{0pt}{0pt}
		$[\delta_{CP}^\ell,\ \alpha_{21},\,\alpha_{31}]$ &  $[200^\circ,\, 349^\circ,\, 291^\circ]$   \\ \hline
		\rule[14pt]{0pt}{0pt}
		$\sum m_i$ &  $70$\,meV      \\ \hline
		\rule[14pt]{0pt}{0pt}
		$\langle m_{ee} \rangle$ &  $8.58$\,meV      \\ \hline
		\rule[14pt]{0pt}{0pt}
		$\epsilon_1$ &  $5.43\times10^{-8}$     \\ \hline
		\rule[14pt]{0pt}{0pt}
		$Y_B$ &  $8.71\times10^{-11}$     \\ \hline
		\rule[14pt]{0pt}{0pt}
		$\sqrt{\Delta\chi^2}$ &  $1.90$     \\ \hline
		\hline
	\end{tabular}
	\caption{Numerical benchmark point of our input parameters and observables at nearby the fixed point $\tau= i \times \infty$ in NH. Here, we  take BP so that $\delta_{CP}$ is closest to the BF value of $195^\circ$ within $\sqrt{\Delta \chi^2}\le1$.}
	\label{bp-tab_nh}
\end{table}
%

\section{Conclusion and discussion}
\label{sec:conclusion}

{We explore  a modular $A_4$ invariant radiative seesaw model.
In this model, we can generate the measured baryon asymmetry via resonant leptogenesis as well as the observed neutrino oscillation data.
Larger Im[$\tau$] leads to the more degenerate Majorana masses.  
This degenerate masses are preferred by the resonant leptogenesis since larger Im[$\tau$]  can enhance the CP asymmetry and avoid the washout processes even though keeping the small Yukawa couplings.  
}
While the nonzero Re[$\tau$] provides the source of CP asymmetry in addition to free complex parameters. In our case, especially, Re[$\tau$] is main source of the CP asymmetry by our $\chi^2$ numerical analysis that would be interesting result in this scenario. Since the constraint of BAU in Eq.~(\ref{bau}) is very strict, we have obtained narrow allowed region as we have discussed in the numerical analysis.
Below, we have highlightend several remarks on the neutrino sector, 
 \begin{enumerate}
\item
We have obtained two small localized regions of {
$150^\circ \lesssim \delta_{\rm CP}\lesssim 170^\circ$ and $180^\circ \lesssim \delta_{\rm CP}\lesssim 200^\circ$ with $65\ {\rm meV}\lesssim \sum m_i \lesssim 80\ {\rm meV}$},
where $180^\circ \lesssim \delta_{\rm CP}\lesssim 200^\circ$ corresponds to the region of ${\rm Re}[\tau]<0$ while $150^\circ \lesssim \delta_{\rm CP}\lesssim 170^\circ$ corresponds to the region of ${\rm Re}[\tau]>0$.
{\it Our result would favor  the best fit value of $\delta_{\rm CP}=195^\circ$.}
\item
We have found two localized islands of {$10^\circ\lesssim\alpha_{21}\lesssim30^\circ,\ 55^\circ\lesssim\alpha_{31}\lesssim80^\circ$,
 and $330^\circ\lesssim\alpha_{21}\lesssim360^\circ,\ 280^\circ\lesssim\alpha_{31}\lesssim310^\circ$,}
where $330^\circ\lesssim\alpha_{21}\lesssim360^\circ,\ 280^\circ\lesssim\alpha_{31}\lesssim310^\circ$ corresponds to the region of ${\rm Re}[\tau]<0$ while $10^\circ\lesssim\alpha_{21}\lesssim30^\circ,\ 55^\circ\lesssim\alpha_{31}\lesssim80^\circ$ one corresponds to the region of ${\rm Re}[\tau]>0$.
 \item
 We have gotten  $7 \, {\rm meV}\lesssim \langle m_{ee}\rangle\lesssim12 \, {\rm meV}$
 and $6 \, {\rm meV}\lesssim  m_{1}\lesssim13 \, {\rm meV}$.
 \end{enumerate}
Furthermore, our heavy neutrino masses and inert scalar bosons are order of several hundred GeV to $\mathcal{O}$(TeV) scale. 
We thus expect that our model can be tested at collider experiments such as the LHC. 
For example, inert scalar bosons from doublet can be produced via electroweak interactions. 
Detailed collider analysis is beyond the scope of this paper and it will be done in future work.

\section*{Acknowledgments}
\vspace{0.5cm}
This work is in part supported by KIAS Individual Grants, Grant  No. PG074202 (JK) and Grant  No. PG076202 (DWK) at Korea Institute for Advanced
Study. 
This research was supported by an appointment to the JRG Program at the APCTP through the Science and Technology Promotion Fund and Lottery Fund of the Korean Government. This was also supported by the Korean Local Governments - Gyeongsangbuk-do Province and Pohang City (H.O.). 
H. O. is sincerely grateful for the KIAS member, and log cabin at POSTECH to provide nice space to come up with this project. 
The work was also supported by the Fundamental Research Funds for the Central Universities (T.~N.).

\appendix

\section{Formulas in modular $A_4$ framework}

Here we summarize some formulas of $A_4$ modular symmetry framework. 
Modular forms are 
holomorphic functions of modulus $\tau$, $f(\tau)$, which are transformed by
\begin{align}
& \tau \longrightarrow \gamma\tau= \frac{a\tau + b}{c \tau + d}\ ,~~ {\rm where}~~ a,b,c,d \in \mathbb{Z}~~ {\rm and }~~ ad-bc=1,  ~~ {\rm Im} [\tau]>0 ~, \\
& f(\gamma\tau)= (c\tau+d)^k f(\tau)~, ~~ \gamma \in \Gamma(N)~ ,
\end{align}
where $k$ is the so-called as the  modular weight.

A superfield $\phi^{(I)}$  is transformed under the modular transformation as 
\begin{equation}
\phi^{(I)} \to (c\tau+d)^{-k_I}\rho^{(I)}(\gamma)\phi^{(I)},
\end{equation}
where  $-k_I$ is the modular weight and $\rho^{(I)}(\gamma)$ represents an unitary representation matrix  corresponding to $A_4$ transformation.
Thus superpotential is invariant if sum of modular weight from fields and modular form in corresponding term is zero (also it should be invariant under $A_4$ and gauge symmetry).

The basis of modular forms is weight 2, $ Y^{(2)}_3=  (y_{1},y_{2},y_{3})$,  transforming
as a triplet of $A_4$ that is written in terms of the Dedekind eta-function $\eta(\tau)$ and its derivative \cite{Feruglio:2017spp}:
\begin{eqnarray} 
\label{eq:Y-A4}
y_{1}(\tau) &=& \frac{i}{2\pi}\left( \frac{\eta'(\tau/3)}{\eta(\tau/3)}  +\frac{\eta'((\tau +1)/3)}{\eta((\tau+1)/3)}  
+\frac{\eta'((\tau +2)/3)}{\eta((\tau+2)/3)} - \frac{27\eta'(3\tau)}{\eta(3\tau)}  \right), \nonumber \\
y_{2}(\tau) &=& \frac{-i}{\pi}\left( \frac{\eta'(\tau/3)}{\eta(\tau/3)}  +\omega^2\frac{\eta'((\tau +1)/3)}{\eta((\tau+1)/3)}  
+\omega \frac{\eta'((\tau +2)/3)}{\eta((\tau+2)/3)}  \right) , \label{eq:Yi} \\ 
y_{3}(\tau) &=& \frac{-i}{\pi}\left( \frac{\eta'(\tau/3)}{\eta(\tau/3)}  +\omega\frac{\eta'((\tau +1)/3)}{\eta((\tau+1)/3)}  
+\omega^2 \frac{\eta'((\tau +2)/3)}{\eta((\tau+2)/3)}  \right)\,, \nonumber \\
 \eta(\tau) &=& q^{1/24}\Pi_{n=1}^\infty (1-q^n), \quad q=e^{2\pi i \tau}, \quad \omega=e^{2\pi i /3}.
\nonumber
\end{eqnarray}
%
Modular forms with higher weight can be obtained from $y_{1,2,3}(\tau)$ through the $A_4$ multiplication rules as shown the last part.
Thus, some $A_4$ triplet modular forms used in our analysis are derived as follows: 
\begin{align}
Y^{(4)}_3&\equiv (y^{(4)}_1,y^{(4)}_2,y^{(4)}_3) = ( y^2_1- y_2 y_3, y_3^2- y_1y_2,y_2^2- y_1y_3),\\
Y^{(6)}_3&\equiv (y^{(6)}_1,y^{(6)}_2,y^{(6)}_3) = ( y^3_1+2y_1 y_2 y_3, y_1^2y_2+2 y^2_2 y_3, y^2_1 y_3+2 y^2_3 y_2),\\
Y^{(6)}_{3'}&\equiv (y^{'(6)}_1,y^{'(6)}_2,y^{'(6)}_3) = ( y^3_3+2y_1 y_2 y_3, y^2_3 y_1+2 y^2_1 y_2, y^2_3 y_2+2 y^2_2 y_1).
\end{align}
$A_4$ multiplication rules are given by 
\begin{align}
\begin{pmatrix}
a_1\\
a_2\\
a_3
\end{pmatrix}_{\bf 3}
\otimes 
\begin{pmatrix}
b_1\\
b_2\\
b_3
\end{pmatrix}_{\bf 3'}
&=\left (a_1b_1+a_2b_3+a_3b_2\right )_{\bf 1} 
\oplus \left (a_3b_3+a_1b_2+a_2b_1\right )_{{\bf 1}'} \nonumber \\
& \oplus \left (a_2b_2+a_1b_3+a_3b_1\right )_{{\bf 1}''} \nonumber \\
&\oplus \frac13
\begin{pmatrix}
2a_1b_1-a_2b_3-a_3b_2 \\
2a_3b_3-a_1b_2-a_2b_1 \\
2a_2b_2-a_1b_3-a_3b_1
\end{pmatrix}_{{\bf 3}}
\oplus \frac12
\begin{pmatrix}
a_2b_3-a_3b_2 \\
a_1b_2-a_2b_1 \\
a_3b_1-a_1b_3
\end{pmatrix}_{{\bf 3'}\  } \ , \nonumber \\
a_{1'}\otimes 
\begin{pmatrix}
b_1\\
b_2\\
b_3
\end{pmatrix}_{\bf 3}
&=a\begin{pmatrix}
b_3\\
b_1\\
b_2
\end{pmatrix}_{\bf 3},\quad
a_{1''}\otimes 
\begin{pmatrix}
b_1\\
b_2\\
b_3
\end{pmatrix}_{\bf 3}
=a\begin{pmatrix}
b_2\\
b_3\\
b_1
\end{pmatrix}_{\bf 3},
\nn\\
{\bf 1} \otimes {\bf 1} = {\bf 1} \ , \quad &
{\bf 1'} \otimes {\bf 1'} = {\bf 1''} \ , \quad
{\bf 1''} \otimes {\bf 1''} = {\bf 1'} \ , \quad
{\bf 1'} \otimes {\bf 1''} = {\bf 1} \ .
\end{align}

\end{document}